\documentclass[conference]{IEEEtran}
\IEEEoverridecommandlockouts

\usepackage{times}
\usepackage{epsfig}
\usepackage{tabularx}
\usepackage{url}
\usepackage{todonotes}
\usepackage{soul}
\usepackage{wasysym}
\usepackage{pgfplots}
\pgfplotsset{compat=newest}

\usepackage{enumitem}
\usepackage{array}
\usepackage{multirow}
\usepackage{rotating}
\usepackage{caption}
\usepackage{subcaption}
\usepackage{adjustbox}
\usepackage{cite}
\usepackage{amssymb,amsfonts}
\usepackage{algorithmic}
\usepackage{graphicx}
\usepackage{textcomp}
\usepackage[super]{nth}
\usepackage{xcolor}
\def\BibTeX{{\rm B\kern-.05em{\sc i\kern-.025em b}\kern-.08em
    T\kern-.1667em\lower.7ex\hbox{E}\kern-.125emX}}
\begin{document}

\title{Blockchain in the Internet of Things:\\ Architectures and Implementation
}


\author{Oscar Delgado-Mohatar, Ruben Tolosana, Julian Fierrez and Aythami Morales \\
School of Engineering, Universidad Autonoma de Madrid, Spain \\
{\tt\small \{oscar.delgado, ruben.tolosana, julian.fierrez, aythami.morales\}@uam.es}
}


\maketitle

\begin{abstract}
The world is becoming more interconnected every day. With the high technological evolution and the increasing deployment of it in our society, scenarios based on the Internet of Things (IoT) can be considered a reality nowadays. However, and before some predictions become true (around 75 billion devices are expected to be interconnected in the next few years), many efforts must be carried out in terms of scalability and security.   

In this study we propose and evaluate a new approach based on the incorporation of Blockchain into current IoT scenarios. The main contributions of this study are as follows: \textit{i)} an in-depth analysis of the different possibilities for the integration of Blockchain into IoT scenarios, focusing on the limited processing capabilities and storage space of most IoT devices, and the economic cost and performance of current Blockchain technologies; \textit{ii)} a new method based on a novel module named BIoT Gateway that allows both unidirectional and bidirectional communications with IoT devices on real scenarios, allowing to exchange any kind of data; and \textit{iii)} the proposed method has been fully implemented and validated on two different real-life IoT scenarios, extracting very interesting findings in terms of economic cost and execution time. The source code of our implementation is publicly available in the Ethereum testnet.

\end{abstract}


\section{Introduction}

The Internet of Things (IoT) has the potential to change the world, just as the Internet did~\cite{ashton2009internet}. This term is referred to the set of objects, sensors, and everyday items that are equipped with computing capability and network connectivity to send/receive data through the Internet\cite{atzori2010internet}. As a result, IoT devices can generate and manage an autonomous ecosystem without any human intervention or supervision. 

Scenarios based on the IoT can already be considered a reality nowadays, for example: \textit{i)} smart homes where the electric light, heating, and kitchen equipment such as the fridge or washing machine are automatically operating, reporting continuosly to the User/Client, and \textit{ii)} autonomous electric vehicles searching for a charging station so that as soon as the car is running out of battery, it automatically drives to the cheapest or nearest point, and starts the charging process. Once completed, the car conducts the payment\cite{Wachenfeld2016}. These are just some of the many applications of the IoT. With the increasing evolution and deployment of the technology in our lives, it is estimated that between 50 and 75 billion devices will be interconnected by 2025 \cite{statista2020,ceo_ericcson}.

Undoubtedly, and before this comes true, a lot of efforts must be carried out in order to manage such volume of information in a scalable and secure way. Many recent studies have focused on the IoT security~\cite{atzori2010internet}. Additionally, key aspects of these low-cost IoT devices such as the limited processing capability and storage space must be further studied~\cite{Morchon2019}. Also, most IoT devices do not usually include protection against physical attacks, so they can be compromised easily. This is exacerbated by the fact that IoT devices almost never have maintenance/upgrade capabilities to reduce production costs. 

In order to mitigate these problems, different approaches have been proposed in the literature such as lightweight cryptography~\cite{Lee2014, Dutta2019}, reinforcement of the perimeter security through the use of firewalls~\cite{Gupta2017, Sari2019}, and zero-trust approaches~\cite{Samaniego2018, Luca2018}. Furthermore, recent studies have put their eyes on other technologies such as Blockchain to overcome some of the limitations existing in IoT scenarios~\cite{dai2019blockchain}. 

Blockchain is essentially a decentralized public ledger of all data and transactions that have ever been executed in the system \cite{swan2015blockchain}. These transactions are recorded in blocks that are created and added to the Blockchain in a linear, chronological order (immutable). Each participating node in the network has the task of relaying transactions, and has a copy of the Blockchain. Other nodes, called miners, are also in charge of validating transactions, performing an expensive computational process, for which they are economically rewarded.

Blockchain was originally created and applied as an auxiliar technology of Bitcoin~\cite{Nakamoto}, providing a secure record of the economic transactions between users of the system. Nevertheless, a Blockchain could store any kind of digital information, providing its certification and guaranteeing its authenticity and integrity. As a result, from its origin up to now, Blockchain has been deployed in many different scenarios such as: biometrics, certification of documentation, mortgages, securities and any other official documents, assets and intelligent objects that can make decisions based on the information stored in the Blockchain, distributed markets without central authority, deposit and custody services that can resolve disputes between customers and merchants, savings accounts, voting systems, and improvements in the distribution chain for all kind of products~\cite{swan2015blockchain,REYNA2018173,2019_ICBA_BlockchainBiometrics,buchmann2017enhancing}.  

\begin{figure*}[t]
\centering
\includegraphics[width=\textwidth]{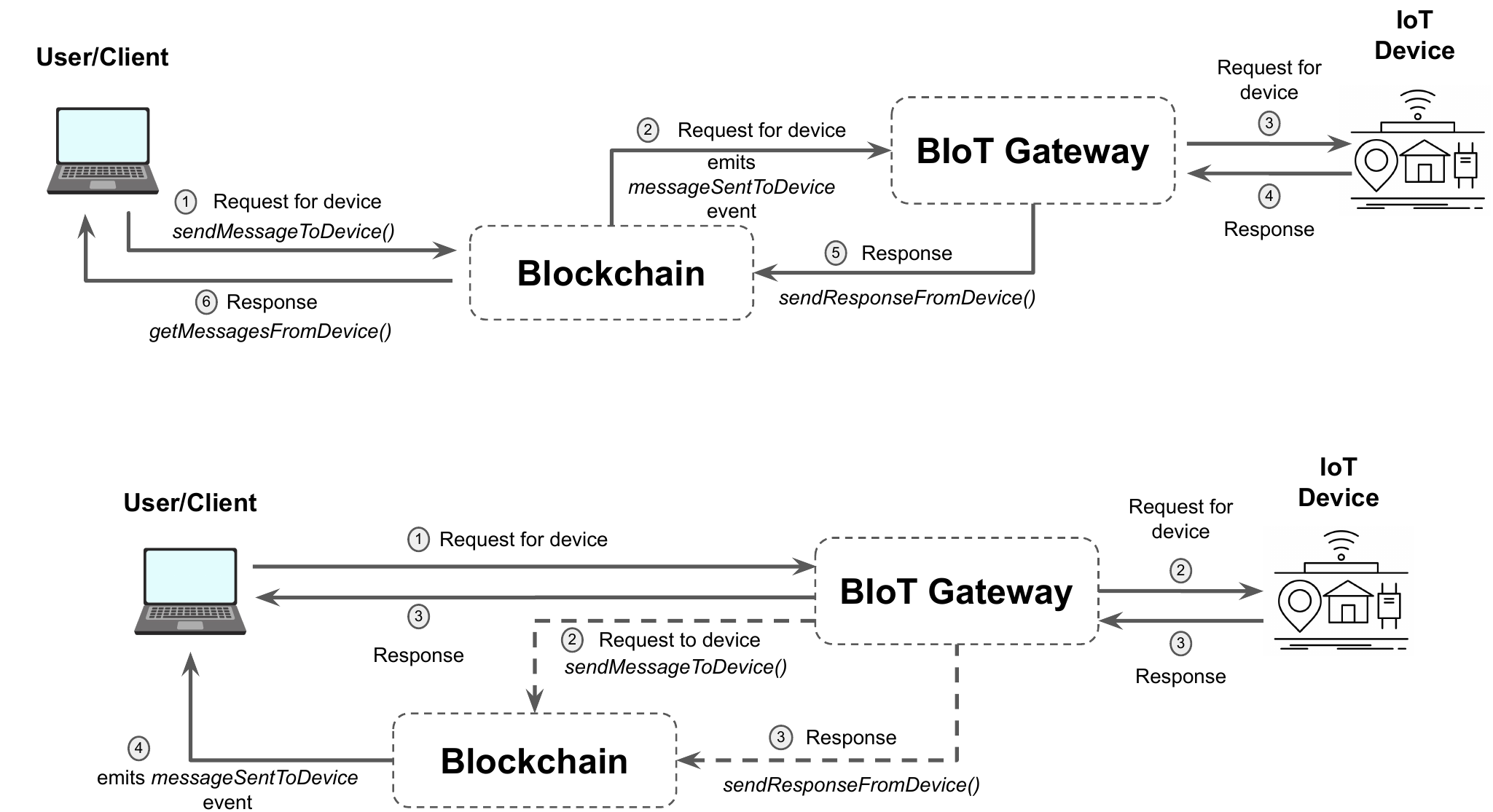}
\caption{Description of our two proposed architectures to incorporate Blockchain into IoT scenarios. Two different configurations are evaluated regarding which module is reached first from the User/Client: (top) Client-Blockchain-Gateway (CBG), and (bottom) Client-Gateway-Blockchain (CGB). The functions and events implemented in our smart contract are detailed in Table~\ref{tab:smart-contract-functions} for a complete understanding of the framework. Dashed lines indicate optional operations.}
\label{fig:configurations}
\end{figure*}

However, and despite these opportunities, the current Blockchain technology suffers from some potential limitations that must be carefully studied and characterized before the adequate integration of Blockchain into IoT scenarios. 

In this study we propose and evaluate a new method for incorporating Blockchain into current IoT scenarios. As a first approximation, Blockchain technology could provide IoT scenarios with some desirable  properties such as immutability, accountability, availability, and universal access. These properties enabled by Blockchain may be very useful for the IoT, among other things, to improve security~\cite{Huynh2019}, transaction reliability~\cite{Gervais2018}, transparency~\cite{REYNA2018173} or privacy levels~\cite{Gervais2018}.

The main contributions of this study can be summarized as follows:
\begin{itemize}
\item An in-depth analysis of the different possibilities for the integration of Blockchain into current IoT scenarios, focusing on: \textit{i)} the limited processing capabilities and storage space of most IoT devices, and  \textit{ii)} the economic cost and performance of using a Blockchain.
\item A novel method based on a new module named BIoT Gateway that allows both unidirectional and bidirectional communications with IoT devices on real scenarios, allowing to exchange any kind of data. 
\item The proposed method has been implemented and validated on two different real-life IoT scenarios, extracting very interesting findings in terms of economic cost and execution time. The source code of our implementation is publicly available in the Ethereum testnet\footnote{0x89f04bFE1c8dbbdbA7c2A7b7815A4A3b229989f8}, and can be verified using explorers such as Etherscan.
\end{itemize}

The remainder of the paper is organized as follows. Sec.~\ref{proposedApproach} describes our proposed methods for the incorporation of Blockchain into IoT scenarios. Sec.~\ref{sec:experimentalSetup} describes all details of our experimental setup in order to validate our proposed approach on practical scenarios. Sec.~\ref{experimentalResults} describes the experimental results achieved using our proposed approach. Finally, Sec.~\ref{conclusions} draws the final conclusions and points out some future research lines.

\begin{table*}[t]
\centering
\caption{Description of all the functions implemented in our smart contract to perform read and store operations. Some functions can be called only by the administrator of the platform ($\CIRCLE$), and others only by the authorized IoT Device ($\LEFTcircle$).}
\label{tab:smart-contract-functions}
\resizebox{\textwidth}{!}{%
\begin{tabular}{|c|c|c|c|c|c|}
\hline
\multicolumn{1}{|c|}{\textbf{Function}} & \textbf{Caller} & \multicolumn{1}{c|}{\textbf{\begin{tabular}[c]{@{}c@{}}Input \\ arguments\end{tabular}}} & \multicolumn{1}{c|}{\textbf{\begin{tabular}[c]{@{}c@{}}Output \\ arguments\end{tabular}}} & \multicolumn{1}{c|}{\textbf{Restriction}} & \textbf{Description} \\ \hline
\textit{registerGateway()} & Administrator & gatewayAddr & - & $\CIRCLE$ & \begin{tabular}[c]{@{}c@{}}It registers a new BIoT Gateway in the system, using its Ethereum\\ address (gatewayAddr). Only the system Administrator can call this function.\end{tabular} \\ \hline
\textit{registerDevice()} & Administrator & \begin{tabular}[c]{@{}c@{}}deviceID, \\ gatewayAddr\end{tabular} & - & $\CIRCLE$ & \begin{tabular}[c]{@{}c@{}}It registers a new IoT Device (deviceID) for a specific gateway (gatewayAddr). \\ Messages from other IoT devices will be discarded. Only the system \\ Administrator can call this function.\end{tabular} \\ \hline
\textit{sendMessageToDevice()} & User/Client & \begin{tabular}[c]{@{}c@{}}deviceID, \\ message\end{tabular} & - & $\LEFTcircle$ & \begin{tabular}[c]{@{}c@{}}It sends a message to the IoT Device (deviceID). It also emits \\ an event \textit{messageSentToDevice} that can be captured by the \\ BIoT Gateway or the Client App.\end{tabular} \\ \hline
\textit{sendResponseFromDevice()} & \begin{tabular}[c]{@{}c@{}}BIoT  Gateway\end{tabular} & \begin{tabular}[c]{@{}c@{}}deviceID, \\ message\end{tabular} & - & $\LEFTcircle$ & \begin{tabular}[c]{@{}c@{}}It stores the response from the IoT Device (deviceID). It also\\ emits an event \textit{responseSentFromDevice}.\end{tabular} \\ \hline
\textit{getMessagesFromDevice()} & User/Client & deviceID & \begin{tabular}[c]{@{}c@{}}Array with \\ messages \\ generated from \\ device (deviceID)\end{tabular} & $\LEFTcircle$ & \begin{tabular}[c]{@{}c@{}}It retrieves all messages generated by the device (deviceID) \\ since the last time this function was called.\end{tabular} \\ \hline
\end{tabular}%
}
\end{table*}

\section{Proposed Methods}\label{proposedApproach}
Fig.~\ref{fig:configurations} shows a graphical representation of our two proposed architectures. They comprise four main modules: User/Client, Blockchain, our proposed BIoT Gateway, and the IoT Device. These two proposed architectures differ in which module (Blockchain or BIoT Gateway) is reached first from the User/Client: \textit{i)} Client-Blockchain-Gateway (CBG), and \textit{ii)} Client-Gateway-Blockchain (CGB). It is important to highlight that our proposed architectures allow unidirectional and bidirectional communications between both the User/Client and the IoT Device, expanding their use to many different practical scenarios. We now describe in detail each module involved in the architecture and also each proposed configuration.

\subsection{Modules}\label{architecture}

\subsubsection{\textbf{User/Client}}\label{client}
It can take any form, typically a web or mobile application that serves as the interface with the final user, e.g., to turn on/off a smart light. This Client App is typically in charge of sending commands to the IoT devices, and receiving responses or readings from them. In our proposed architectures, the User/Client App can communicate with the IoT Device through both the BIoT Gateway or the Blockchain, depending on the selected configuration (Sec.~\ref{configurations}).

\subsubsection{\textbf{Blockchain}}\label{Blockchain}
We assume a Blockchain capable of storing data and running code through smart contracts, a well-known concept inside the cryptographic community \cite{Szabo1996}. A smart contract is, essentially, a piece of code executed in a secure environment that controls digital assets. This concept has not been popular until its inclusion in the Ethereum Blockchain platform~\cite{Dannen2017}. In essence, Ethereum could be seen as a distributed computer, with capability to execute programs written in Turing-complete, high-level programming languages.

In our proposed method, we have developed a smart contract in order to enable a secure two-way communication, in which the integrity of the exchanged data is guaranteed by the smart contract and the underlying Blockchain. As a result, we keep track of the incoming and outgoing messages between the User/Client and the IoT Device, and also control the devices registered in the BIoT Gateway to prevent unauthorized use. 

Essentially, two operations are implemented in the smart contract to read and store the data coming from or going to the IoT Device. Data can have any meaning, such as control commands, readings from the sensors or any other type of digital data. All the supported functions needed to perform the read and store operations are described in Table \ref{tab:smart-contract-functions}, including a description of each of them, input/output arguments, permission restriction, and which module should call them.

\subsubsection{\textbf{BIoT Gateway}}\label{gateway}
One of the main modules of our proposed architectures is the novel BIoT Gateway. It acts as an interface between all the modules of the architecture, allowing three possible communications: \textit{i)} the interface with the User/Client, usually through an API REST (web/mobile application); \textit{ii)} the interface with the Blockchain (smart contract); and \textit{iii)} the interface with the IoT Device module, for example, a MQTT broker or any other protocol such as HTTPS, as long as it provides a secure communication, and the authentication of both ends. It is important to remark that the inclusion of the BIoT Gateway breaks the intrinsic distributed nature of the Blockchain. Nevertheless, this approach remains valid and secure in most scenarios. In fact, if the same entity manages both the BIoT Gateway and the IoT Device, they form a logical unit from the point of view of trust, so the security obtained would be equivalent to place the Client directly on the Device.

\subsubsection{\textbf{IoT Device}}\label{IoT_device}
The last module of the proposed architecture is the IoT Device. It is important to remark that no especial hardware or software capabilities of the IoT Device are needed when considering our proposed method, making it feasible for any low-cost device. This is one of the main advantages of our proposed methods.

\subsection{Configurations}\label{configurations}

\subsubsection{\textbf{Client-Blockchain-Gateway (CBG)}}\label{CBG}
The User/Client first communicates with the Blockchain module, which broadcasts an event that is captured by the BIoT Gateway, and then moved forward to the IoT Device. Later on, the command/data is sent back to the BIoT Gateway from the IoT Device, storing the command/data into the Blockchain. Finally, it is sent back to the User/Client. 

This is the most secure configuration as the User/Client can always prove that one message has been sent, even if the BIoT Gateway refuses to process it (non-repudiation). In return, a considerable delay can be introduced under some circumstances. Therefore, this configuration is more suitable for scenarios without strong latency requirements such as in the certification of information.

\subsubsection{\textbf{Client-Gateway-Blockchain (CGB)}}\label{CGB}
In this case, the communication between the User/Client and the IoT Device occurs first through the BIoT Gateway. Once the command/data is received by the BIoT Gateway, it is moved forward to the IoT Device, and optionally, to the Blockchain to assure its integrity as soon as possible. Later on, the command/data is first sent back to the BIoT Gateway and then to the User/Client and optionally, again to the Blockchain, to assure its integrity as soon as possible. 

This approach reduces the overhead due to the optional use of the Blockchain, but in return, security can be affected. Therefore, this configuration should only be considered in scenarios where real-time communications are needed, such as a smart-home scenario as the security of the messages exchanged is in theory not critical.

\section{Experimental Setup}\label{sec:experimentalSetup}

\subsection{IoT Scenarios}\label{IoTScenarios}
Our proposed method has been implemented and validated in a real environment. Two different IoT scenarios are recreated, considering both unidirectional and bidirectional configurations:

\begin{itemize}
    \item \textbf{Refrigerated container}: the IoT Device periodically sends its temperature and other metadata (24bytes) to the BIoT Gateway, once per minute. A unidirectional communication is considered in this scenario.
    \item \textbf{Smart light}: the IoT Device can receive and send simple commands (20 per day) to turn on/off the light (24 bytes), simulated with a LED matrix. A bidirectional communication is considered in this scenario.
\end{itemize} 

\subsection{Implementation Details}\label{ImplementationDetails}
For the implementation of each scenario, the following details are considered in our experiments: 

\begin{itemize}
\item \textbf{User/Client}: a personal computer is used to receive and send commands to the IoT Device.
\item \textbf{Blockchain}: a smart contract\footnote{It is a basic contract that should be considered only for research purposes.} has been developed in Solidity language, and deployed in Ethereum Ropsten testnet\footnote{0x89f04bFE1c8dbbdbA7c2A7b7815A4A3b229989f8}. This platform is functionally identical to the main platform, but allows development and testing of applications without economic cost.  
\item \textbf{BIoT Gateway}: it has been implemented using a Raspberry Pi 4~\cite{RaspberryPi4}, running the official Ethereum client (Geth), and connected to the Ropsten testnet in light mode.
\item \textbf{IoT Device}: we consider the Wemos D1 mini~\cite{wemos}, a popular low-cost microcontroller based on the ESP8266 platform, connected to a temperature sensor and a LED matrix. All the elements are operated with a battery.
\end{itemize}

In our laboratory setup, both the BIoT Gateway and the IoT Device are connected through the MQTT protocol, although, as stated before, any other secure protocol may be used. Of course, this protocol must allow the mutual authentication of both elements, because they form a single ``logical unit'' in terms of trust. In our case, both BIoT Gateway and IoT Device are issued with a x509 certificate by a common CA. The fingerprint of the certificate of the BIoT Gateway is hard-coded (certificate pinning), to avoid man-in-the-middle (MITM) attacks.

\subsection{Blockchain Storage Schemes}\label{BlockchainStorage}
One of the main potential limitations for the integration of both IoT and Blockchain technologies is the economic cost of running an IoT system (totally or partially) in Blockchain. It is therefore crucial to properly estimate and minimize the cost that, to a large extent, is due to the storage of data. In our experimental framework we analyze the three different storage schemes proposed in~\cite{delgado2019biometric}, which can be ordered in terms of complexity (from lower to higher), and economic cost (from higher lo lower) as follows:

\begin{itemize}
    \item \textbf{Full on-chain storage}: all data is stored, as-is, in the Blockchain. 
    \item \textbf{Data hashing}: the Blockchain only stores a hash of the data that guarantees its immutability. The data itself is stored off-chain in another system: distributed (e.g., IPFS \cite{Benet2014}), cloud or even existing traditional databases.
    \item \textbf{Merkle trees}: data is also stored off-chain, but it is preprocessed by constructing a Merkle tree structure, which reduces storage costs and increases the bandwidth.
\end{itemize}

These alternatives are discussed in terms of economic cost and execution time in the next section.

\begin{table*}[t]
\centering
\caption{Experimental results in terms of economic cost and performance for each Blockchain storage scheme (i.e., full on-chain, data hashing, and Merkle trees) and architecture configuration (i.e., CBG and CGB). The economic cost and performance results achieved in our two IoT scenarios studied (i.e., refrigerated container and smart light) are included in the last two rows of the table. We have considered a gas price of 1 gwei (1 gwei = $10^{-9}$ ETH), and 1 ETH = \$168 (accurate at time of writing, January 2020). Times have been measured performing each operation ten times, discarding the minimum and maximum times, and calculating the average of the rest. }
\label{tab:results}
\resizebox{\textwidth}{!}{%
\begin{tabular}{|c|c|c|c|c|c|}
\hline
\multirow{2}{*}{\textbf{Configuration}} & \multirow{2}{*}{\textbf{Operation}} & \multicolumn{3}{c|}{\textbf{Economic Cost}} & \textbf{Performance} \\ \cline{3-6} 
 &  & \textit{\begin{tabular}[c]{@{}c@{}}Full on-chain\\ (per message)\end{tabular}} & \textit{\begin{tabular}[c]{@{}c@{}}Data hashing\\ (per message)\end{tabular}} & \textit{\begin{tabular}[c]{@{}c@{}}Merkle trees\\ (for any amount\\ of data)\end{tabular}} & \textit{\begin{tabular}[c]{@{}c@{}}Execution\\ time\\ (average)\end{tabular}} \\ \hline
- & \begin{tabular}[c]{@{}c@{}}Smart contract\\ deployment\end{tabular} & \multicolumn{3}{c|}{\begin{tabular}[c]{@{}c@{}}866,212 gas\\ (\$0.145)\end{tabular}} & 10 secs. \\ \hline
- & \textit{registerGateway()} & \multicolumn{3}{c|}{\begin{tabular}[c]{@{}c@{}}43,702 gas\\ (\$0.007)\end{tabular}} & 15 secs. \\ \hline
\multirow{2}{*}{\begin{tabular}[c]{@{}c@{}}CBG\end{tabular}} & \textit{sendMessageToDevice()} & \begin{tabular}[c]{@{}c@{}}52,132 gas for $m_{size}$ = 16\\ 382,119 gas for $m_{size}$ = 1204\\ \\ (\$0.008 for $m_{size}$=16)\\ (\$0.06 for $m_{size}$=1024)\end{tabular} & \begin{tabular}[c]{@{}c@{}}72,433 gas\\ (\$0.012)\end{tabular} & \begin{tabular}[c]{@{}c@{}}72,433 gas\\ (\$0.012)\end{tabular} & 12 secs. \\ \cline{2-6} 
 & \textit{receiveMessagesFromDevice()} & - & - & - & - \\ \hline
\multirow{2}{*}{\begin{tabular}[c]{@{}c@{}}CGB\end{tabular}} & \textit{sendMessageToDevice()} & \begin{tabular}[c]{@{}c@{}}52,132 gas for $m_{size}$ = 16\\ 382,119 gas for $m_{size}$ = 1204\\ \\ (\$0.008 for $m_{size}$=16)\\ (\$0.06 for $m_{size}$=1024)\end{tabular} & \begin{tabular}[c]{@{}c@{}}72,433 gas\\ (\$0.012)\end{tabular} & \begin{tabular}[c]{@{}c@{}}72,433 gas\\ (\$0.012)\end{tabular} & 13 secs. \\ \cline{2-6} 
 & \textit{receiveMessagesFromDevice()} & - & - & - & - \\ \hline
\textbf{Scenario} & \textbf{Operation} & \multicolumn{3}{c|}{\textit{\begin{tabular}[c]{@{}c@{}}(per 24h of operation and device)\\ (8 data bytes + 16 device ID bytes)\end{tabular}}} & \begin{tabular}[c]{@{}c@{}}(Per \\ interaction)\end{tabular} \\ \hline
\begin{tabular}[c]{@{}c@{}}Refrigetared\\ container\end{tabular} & \multicolumn{1}{c|}{\begin{tabular}[c]{@{}c@{}}The temperature is sent every\\ minute to the BIoT Gateway \\ (unidirectional) \end{tabular}} & \$11.52 & \$17.28 & \multirow{2}{*}{\$0.012} & \begin{tabular}[c]{@{}c@{}}BG: 15 secs\\ GB: 0 secs\end{tabular} \\ \cline{1-4} \cline{6-6} 
\begin{tabular}[c]{@{}c@{}}Smart light\end{tabular} & \multicolumn{1}{c|}{\begin{tabular}[c]{@{}c@{}} The light is turned on/off \\20 times a day \\ (bidirectional).\end{tabular}} & \$0.16 & \$0.24 &  & \begin{tabular}[c]{@{}c@{}}BG: 30 secs\\ GB: 0 secs\end{tabular} \\ \hline
\end{tabular}%
}
\end{table*}

\section{Experimental Results}\label{experimentalResults}
Our proposed method has been evaluated according to the following measures:

\begin{itemize}
\item \textbf{Economic Cost}: associated to the use of Blockchain, due to the data storage and smart contract execution. Both CBG and CGB configurations are analyzed. 
\item \textbf{Performance}: associated to the execution time of the smart contract. Transmission times are not included as they are negligible with respect to the smart contract.

\end{itemize}

Table~\ref{tab:results} shows the economic costs and performance of each Blockchain storage scheme (i.e., full on-chain, data hashing, and Merkle trees) and architecture configuration (i.e., CBG and CGB). We also include in the last two rows of the table the economic cost and performance results obtained in our two IoT scenarios studied (i.e., refrigerated container and smart light). The economic cost is described in terms of units of gas and US dollars at the time of writing (January, 2020). Performance is described in terms of seconds.

\subsubsection{\textbf{Economic Cost}}

We first analyze how the different Blockchain storage schemes affect the feasibility of our proposed approach in terms of economic cost. In general, the results depicted in Table~\ref{tab:results} remark that Merkle trees seem to be the only viable Blockchain storage scheme. The remaining storage schemes would quickly become prohibitive for the volume of data typically exchanged in a real environment. We now analyze in detail each Blockchain alternative.

The full on-chain or direct storage scheme is specially expensive (e.g., \$11.52 per day for the refrigerated container scenario) as all data is stored, as-is, in the Blockchain. The reason of this high economic cost is due to the pricing storage in Blockchain, which is intentionally discouraged to minimize its uncontrolled growth. For example, protecting the security of one million messages with this approach would cost between \$8,756 and \$22,565 for messages of 8 and 128 bytes, respectively. Depending on the final scenario and the value of the protected information, this could be reasonable but, in general, these figures are not affordable for general purpose applications. 


The next storage scheme considered is data hashing. This approach slightly improves the economic cost figures compared with the full on-chain storage scheme, but only when the size of the messages is bigger than 32 bytes (for smaller messages the direct storage is still cheaper). Despite the improvement, this storage scheme is still prohibitive in terms of economic cost (e.g., \$17.28 per day for the refrigerated container scenario).

The last storage scheme studied is Merkle trees. In this case, all messages and data received during a period of time are grouped under a single tree. As a result, only the root of the tree must be secured in the Blockchain. Therefore, an arbitrarily large volume of data can be secured at the cost of only 256 bits, having a fixed cost of \$0.0122 per day and IoT Device (in our experimental setup), being this storage scheme the only one viable in real environments. The exact duration of this period (window) of time in which data is grouped under a single tree must be determined taking into account the volume of messages processed, their importance, and the cost that can be assumed. Depending on these parameters, the duration of the window can range from a few minutes to several hours or days. Finally, despite the economic cost advantages of this storage scheme, it is important to remark that if the BIoT Gateway, for whatever reason, is lost or corrupted before the root of the tree can be secured in the Blockchain, then the security of the previous messages is lost. In addition, this scheme slightly complicates the verification of data integrity, since, apart from the message itself, it is necessary to save a cryptographic proof that allows its reconstruction.

Finally, and although the economic cost of this last approach is very low for a single IoT Device (\$0.0122), it could be too high in a realistic IoT environment composed of potentially millions of devices. To mitigate this aspect, a Merkle tree meta-structure could be generated by aggregating the corresponding trees to the individual BIoT Gateways. This way, it would be possible to authenticate and process an arbitrarily large volume of data and messages at a very low fixed cost.

\subsubsection{\textbf{Performance}}
We now analyze how the architecture configuration selected (i.e., CBG and CGB) affects the feasibility of our proposed approach in terms of execution time (last column of Table~\ref{tab:results}). 

In general, the experiments show that this hybrid approach based on our proposed BIoT Gateway is also viable. As can be seen, the execution time is slightly higher than 10 seconds for the \textit{sendMessageToDevice()} operation, which is used in both CBG and CGB configurations. 

This time delay could be acceptable or not depending on the final IoT scenario. For the refrigerated container scenario considered in our experimental setup (unidirectional communication), where the container periodically sends its temperature to the BIoT Gateway, it seems feasible a time delay between 10 and 15 seconds as no hard time constrains are needed in this specific scenario. Therefore, the CBG configuration should be chosen to increase the security.

For the smart light scenario considered (bidirectional communication), in which the IoT Device can receive and send simple commands to turn on/off the light, time delay seems much more sensitive due to usability reasons. Therefore, in this specific scenario, the CGB configuration should be considered as it does not add virtually delays. However, the messages exchanged are not secured in the Blockchain until the end of the window period configured for the system.

Finally, the message retrieval operation, i.e., \textit{receiveMessagesFromDevice()}, is free of charge, as it is a read-only operation and does not include/modify anything of the Blockchain. Furthermore, this operation can be considered immediate in terms of execution time, since the request is processed by the local Ethereum node, and does not reach the network.

\section{Conclusions}\label{conclusions}

In this study we have explored new architectures for incorporating Blockchain into current IoT scenarios. In particular, we have proposed new methods based on a novel module named BIoT Gateway that allows both unidirectional and bidirectional communications with IoT devices on real scenarios, allowing to exchange any kind of data.


Our proposed methods have been fully implemented and validated on two different real-life IoT scenarios: \textit{i)} a refrigerated container with a unidirectional communication, and \textit{ii)} a smart light with a bidirectional communication. Also, three different Blockchain storage schemes are evaluated in order to minimize the economic cost of data storage. 

The results achieved prove that straightforward schemes such as the direct storage of the IoT templates on-chain, or direct data hashing, are not feasible for practical IoT scenarios. Nevertheless, when the Merkle tree scheme is included as an intermediate data structure, the economic cost is significantly reduced and also fixed regardless of the volume of data to store. Regarding the performance, times between 10-20 seconds are obtained for store operations whereas for read operations, they are usually free of cost and very fast to run as they are processed locally. These figures prove the viability of our proposed approach on current IoT scenarios, overcoming some limitations of most IoT devices such as the limited processing capabilities and storage space.

\section*{Acknowledgments}
This work has been supported by projects: PRIMA (H2020-MSCA-ITN-2019-860315), TRESPASS-ETN (H2020-MSCA-ITN-2019-860813), BIBECA (RTI2018-101248-B-I00 MINECO/FEDER), and COPCIS (TIN2017-84844-C2-1-R MINECO/FEDER). Ruben Tolosana is supported by Consejer\'ia de Educaci\'on, Juventud y Deporte de la Comunidad de Madrid y Fondo Social Europeo.

{
\bibliographystyle{IEEEtran}
\bibliography{biblio,blockchain_iot,blockchain_biblio}
}

\end{document}